# COMPENSATION FOR REACTIVE POWER AND HARMONIC CURRENTS DRAWN BY NON-LINEAR LOAD IN PV-MICRO HYDRO GRID

Raj Krishna Nepal
*Dept. of Electrical Engineering*
*Central Campus, Pulchowk, IOE, TU*
Lalitpur, Nepal
075bel034.raj@pcampus.edu.np

Bibek Khanal
*Dept. of Electrical Engineering*
*Central Campus, Pulchowk, IOE, TU*
Lalitpur, Nepal
075bel011.bibek@pcampus.edu.np

Sanket Khatiwada
*Dept. of Electrical Engineering*
*Central Campus, Pulchowk, IOE, TU*
Lalitpur, Nepal

Nirajan Bhandari
*Dept. of Electrical Engineering*
*Central Campus, Pulchowk, IOE, TU*
Lalitpur, Nepal

Bishal Rijal
*Dept. of Electrical Engineering*
*Khwopa College of Engineering, IOE,TU*
Lalitpur, Nepal

Raisha Karmmacharya
Ajay Thapa
*Dept. of Electrical Engineering*
*Central Campus, Pulchowk, IOE, TU*
Lalitpur, Nepal

*Abstract*— This paper presents a simulation approach to enhance the power quality of a PV-micro hydro grid supplying both linear consumer load and non-linear industrial load by integrating Shunt Active Power Filter (SAPF), utilizing instantaneous PQ theory and hysteresis current control band logic. The non-linear load draws reactive power and harmonic current from the source thereby affecting the power quality. The integration of the SAPF at the point of common coupling (PCC) offers reactive power and harmonic current compensation, ensuring that the current supply to the grid remains nearly sinusoidal and proportional to the active power. By injecting equal and opposite harmonic components, the SAPF effectively reduces Total Harmonic Distortion (THD) from 7% to 2.96%, thereby enhancing the overall power quality of the PV-micro hydro grid system.

## I. INTRODUCTION

In context of Nepal, hydro power plants ranging from capacity of 10 kW to 100kW are classified as Micro Hydro Power (MHP) plant[1]. As national grid extension to rural areas with few populations is not economical due to high cost of transmission line and power loss in the line, the concept of MHP plant is very popular in Rural areas of Nepal for electrification. Electronic Load Controller (ELC) is used for speed control instead to conventional speed governor to reduce the cost of plant which dissipates the surplus energy into the dump load and maintain the stability of the system even if the rural areas load is fluctuating in nature.

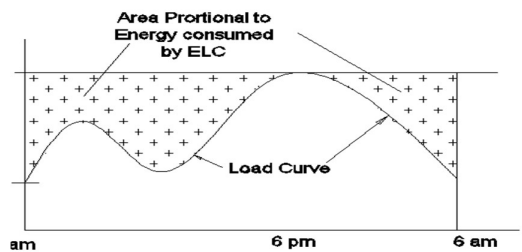

Figure 1: Daily load curve of MHP plant.

Figure 1 shows the daily load curve of MHP plant where the shaded area represents the energy wasted in the dump load. As all the energy supplied by the MHP plant is not fully utilized, the load factor of such plant is very low. The load factor can be increased by encouraging rural industry which has mostly non-linear load during day time. If the sum of maximum demand of rural industry and consumers load at day time is greater than the capacity of existing MHP, PV plant can be added to supply extra power as shown in figure 2.

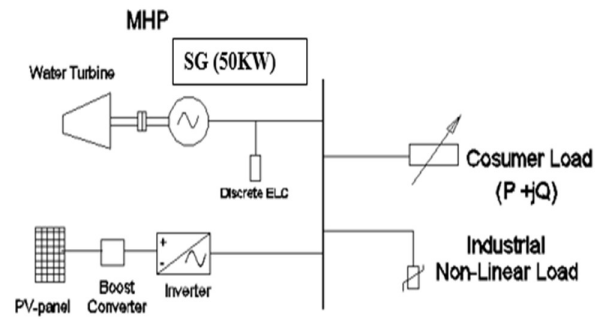

Figure 2: MHP plant with PV source connected to linear consumer load and non-linear industrial load.

The industrial load could be inductive as well as non-linear. Non- linear load draws current in abrupt short pulses. These short pulses distort the current waveform, leading to the generation of harmonics. Harmonics can cause various issues such as poor power factor, overheating, reduced efficiency, and the unnecessary tripping of circuit breakers, all of which contribute to power quality problems [2]. A PV source is unable to support the reactive load and harmonic currents generated by non-linear loads [3]. This issue could become more significant in the future, making it crucial to address it. There are essentially two methods for mitigating power quality issues. The first method is load conditioning, which aims to make the load resistant to harmonics. This involves making equipment less sensitive to harmonics and power disturbances, though it is not always feasible in practice. The second method is power line conditioning. This approach involves installing a line conditioning system at the point of common coupling (PCC) to suppress or counteract the adverse effects produced by non-linear harmonic-generating loads [4]. Traditionally, passive filters were employed to address issues related to harmonic generation and reactive power disturbances. However, they had significant drawbacks, including resonance issues, large size, fixed compensation characteristics, and the impact of source impedance on performance, which made this solution less appealing. As a result, the concept of active power filters was introduced. Active power filters offer a more effective solution compared to conventional passive filters for mitigating harmonic and reactive power disturbance problems [5]. Basically, shunt active power filter operates as a current source injecting the current equal and opposite of harmonic components generated by the load. The use of shunting active filter benefits in reducing the current harmonics produced by non-linear loads and benefits in improving the power factor quality at PCC.

## II. METHODOLOGY

The overall modeling of the system and the observation of the results are done in MATLAB/SIMULINK (2021a). Figure 3 shows the flowchart of the overall methodologies of the system.

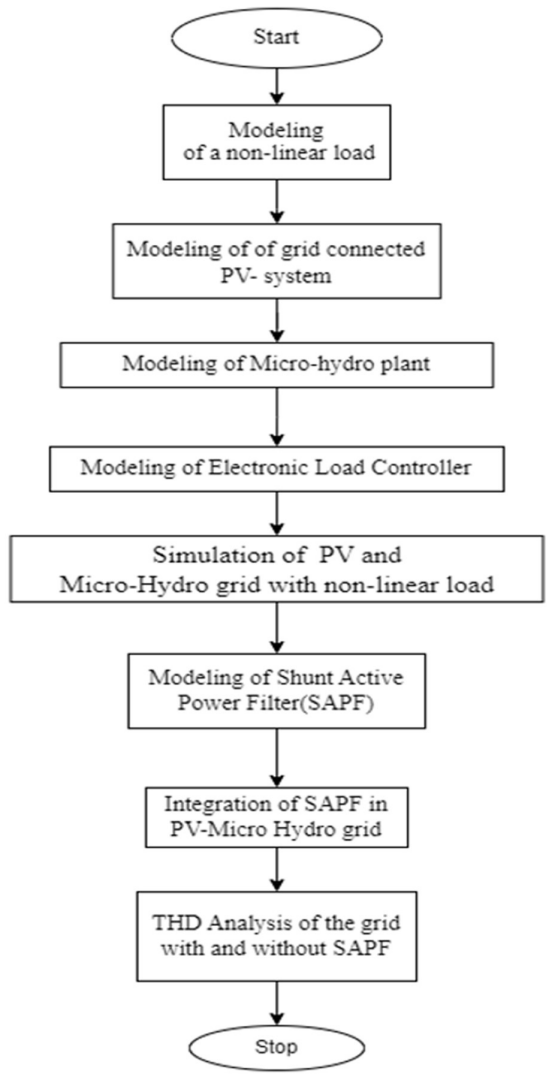

Figure 3: Flowchart showing the overall methodologies of the system.

### A. Modeling of non-linear load

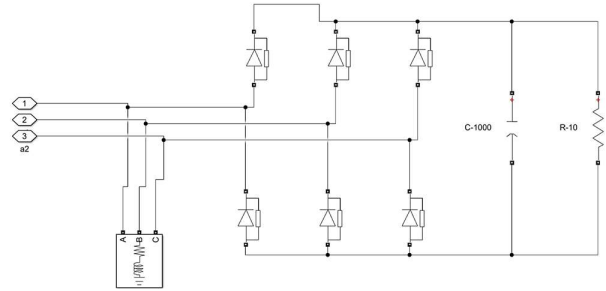

Figure 4: Simulation model of non-linear load.

Figure 4 shows the simulation model of the non-linear system consisting of rectifier circuit that consumes non sinusoidal current. Here we have used diodes as our circuit elements, resistor of 10 ohms and capacitance value of 5000 microfarad.

### B. Modeling of grid connected PV system

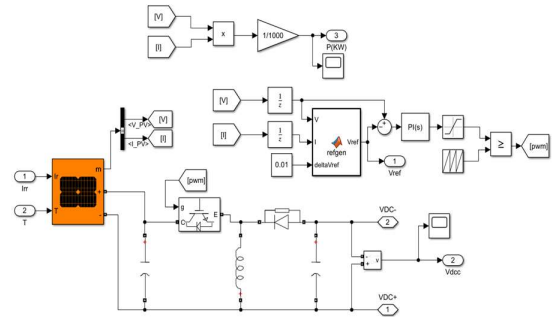

Figure 5: Simulation model of PV system using MPPT.

Figure 5 shows the simulation model of PV system using MPPT algorithm. Here, the 50 KW solar panel is modeled and MPPT is performed by the Perturbation and Observation algorithm to obtain the reference voltage corresponding to the maximum power which is then compared with actual voltage of the solar panel to generate the Pulse Width Modulated (PWM) signal that controls the switching operation of buck boost converter.

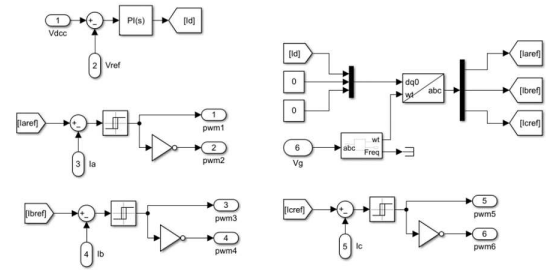

Figure 6: Simulation model of hysteresis band controller.

Figure 6 shows the simulation model of hysteresis band controller which compares the voltage across dc link capacitor with the reference voltage corresponding to the maximum power with the voltage across the dc link capacitor to get the d component of the current and keeping the q component of current to zero, the reference current is generated after the dq to abc conversion.

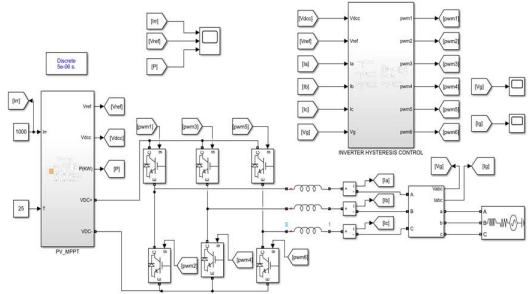

Figure 7: Overall simulation model of grid connected PV – System.

Figure 7 shows the overall simulation model of grid connected PV system using hysteresis band controller. The output of PV_MPPT subsystem from figure 5 i.e., the PV is connected to the three phase voltage source inverter using the buck boost converter and MPPT algorithm and then to the grid through the L filter. The switch operation of the inverter is controlled by the PWM signal generated using the hysteresis band controller from figure 6.

### C. Modeling of Micro-hydro plant

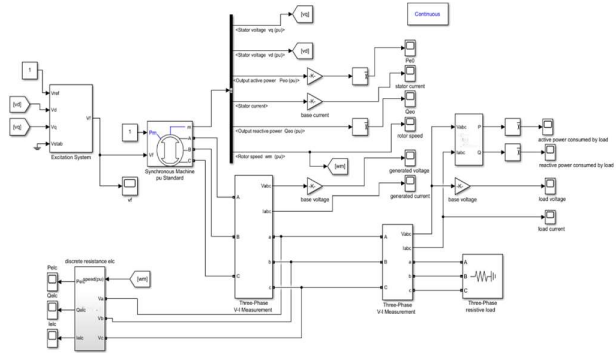

Figure 8: Simulation model of micro-hydro plant.

Figure 8 shows the simulation model of micro-hydro plant. At first the synchronous machine is synchronized controlling the parameters of the excitation system and then connected to the load. The surplus power of the plant during the day time is dissipated in the dump load called discrete resistance ELC (Electronic Load Controller).

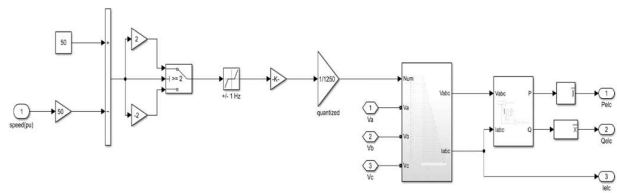

Figure 9: Simulation model of ELC.

Figure 9 shows the simulation model of ELC based on discrete resistance mechanism. In this method the actual rotor speed is compared with the reference speed and then the error signal is quantized. Finally, according to the error in the speed the dump load is turned on to maintain the rotor speed at 1pu according to the error in the speed.

### D. Modeling of Shunt Active Power filter (SAPF)

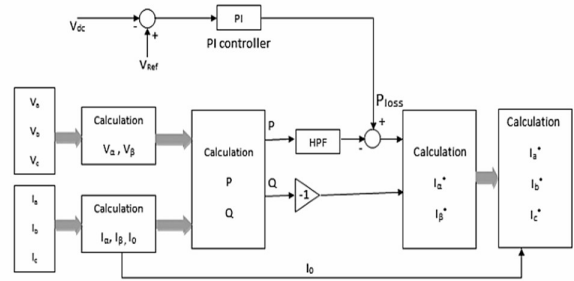

Figure 10: Block diagram of shunt active power filter model.

Figure 10 shows block diagram of shunt active power filter model where the voltage and current from the PV-Micro grid is measured and then Clarke's transformation is performed to get the instantaneous active and reactive power using PQ theory. The dc link voltage is compared to the reference voltage and the internal loss in the compensator is calculated. Then the instantaneous power is passed through a high pass filter and compared with the internal loss to obtain reference current in alpha beta frame. Finally, inverse Clarke transformation is performed to convert alpha beta reference current to abc frame. The reference current is then compared with the actual current to generate the gate pulses for the inverter control using three-phase hysteresis band current controller. And the simulation model of the shunt active power filter is shown in figure 11.

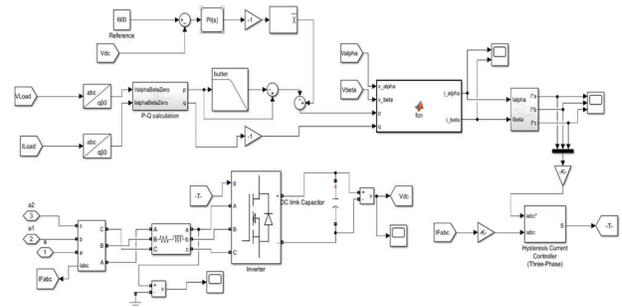

Figure 11: Simulation model of shunt active power filter.

### E. Integration of SAPF in PV-micro hydro grid

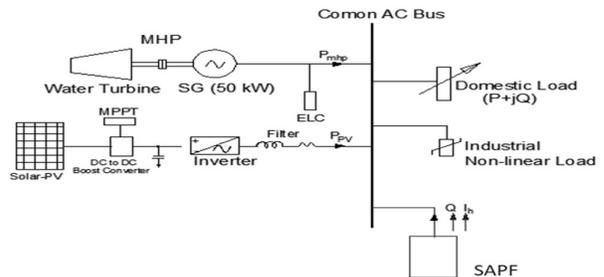

Figure 12: Block diagram of integration of SAPF in PV-micro hydro grid.

Figure 12 shows the overall block diagram of integration of SAPF in PV- micro hydro grid such that the switching operation of inverter used in the SAPF is controlled using hysteresis band controller along with PQ theory to inject current to the AC bus to compensate harmonic current and reactive power drawn by the non-linear load.

## III. SIMULATION RESULTS

### A. Impact of Non-linear load

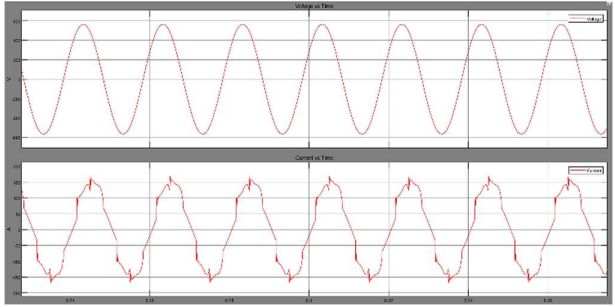

Figure 13: Voltage and current waveform of non-linear load.

Figure 13 depicts the voltage and current waveform of non-linear industrial load which shows that the load draws the non-sinusoidal current from the pure sinusoidal voltage source.

### B. PV system output

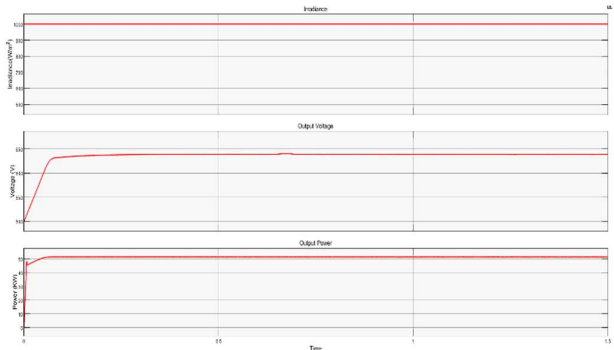

Figure 14: Output voltage and power waveform of PV system.

Figure 14 shows the output voltage and power waveform of the PV system at 1000 W/m$^2$ irradiance. It shows that the PV delivers the maximum power of 50 KW at corresponding voltage of 640V verifying that the PV successfully injects the maximum power to the grid.

### C. MHP system output

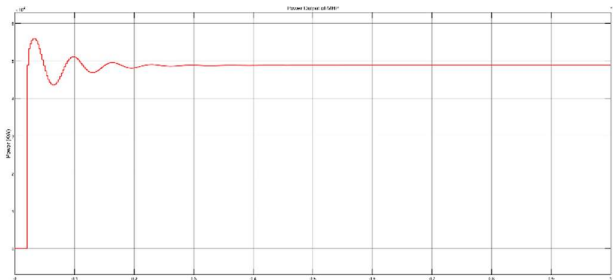

Figure 15: Output power waveform of MHP.

Figure 15 depicts the output power waveform of MHP such that the power generated by the MHP is observed to be approximately 50kw.

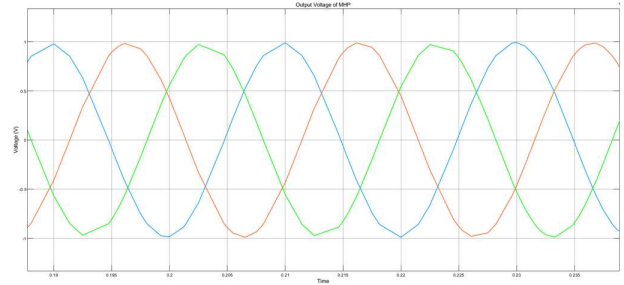

Figure 16: Output voltage waveform of MHP.

Figure 16 shows the output voltage waveform of MHP and indicates that the MHP successfully generates 1pu voltage even at varying load conditions.

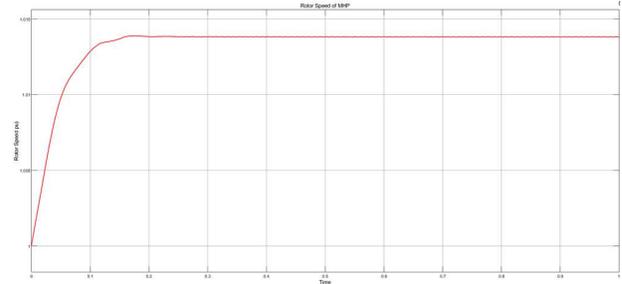

Figure 17: Rotor speed waveform of MHP.

From figure 17, we can observe that the fluctuation in load doesn't affect the rotor speed due to the excess power being dumped in the ELC. In this way the speed control of the MHP is performed successfully at varying load conditions.

### D. Analysis of PV-micro hydro grid without SAPF

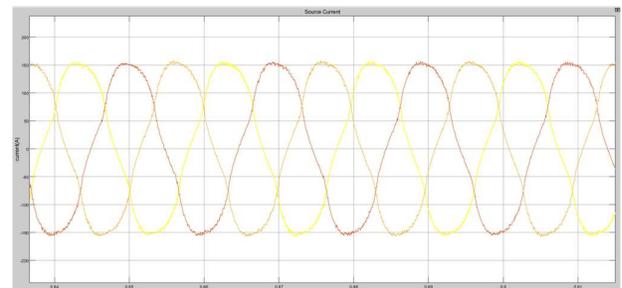

Figure 18: Output current waveform of PV-micro hydro grid without SAPF.

Figure 18 shows the output current waveform of the PV-micro hydro grid without SAPF. Here, we can observe that the source current supplied by the MHP and PV plant is non-linear. This can hamper the windings of the generator used in MHP, as well as the inverter of the PV system.

Total Harmonic Distortion of source current without SAPF is found to be approximately 7% as shown in figure 19. By increasing the non-linear property of the load, it can be varied and can be more in practical load.

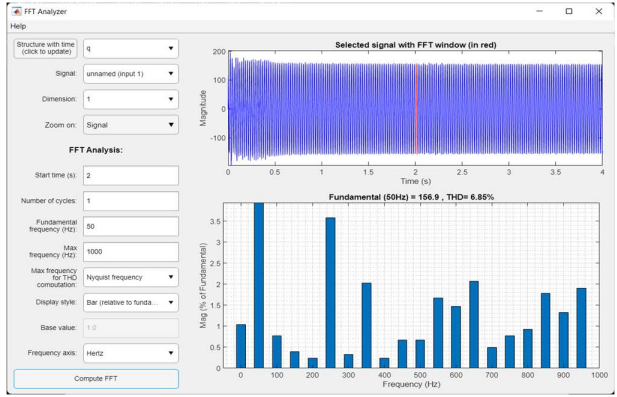

Figure 19: THD analysis of source current without SAPF.

*E. Analysis of PV-micro hydro grid with SAPF*

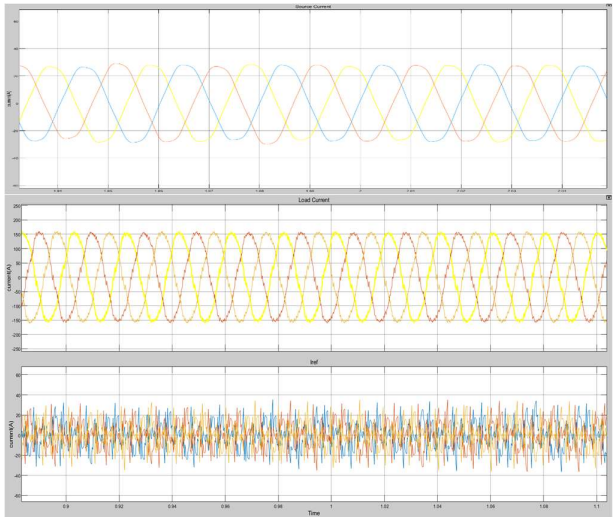

Figure 20: Current waveform of PV-micro hydro grid with SAPF.

Figure 20 depicts the waveform of source current delivered by the PV-micro hydro system, load current drawn by the combination of linear and non-linear industrial load, and the compensating current injected by the SAPF into the system. We can observe that the source is supplying only sinusoidal current whereas the non-linear part of the load current is supplied by the inverter of SAPF.

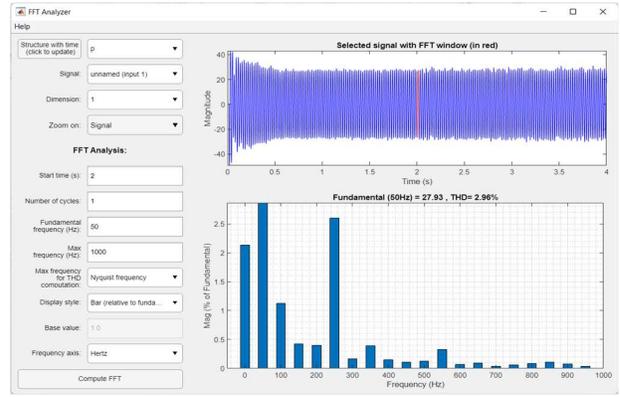

Figure 21: THD analysis of source current with SAPF.

From figure 21, it can be seen that the THD of a source current delivered by the PV-micro hydro is reduced to 2.96% nullifying higher order harmonics after integrating SAPF into the system.

## IV. CONCLUSION

An ELC controlled micro hydro is integrated with the PV system to supply load to both linear and non-linear load. The non-linear industrial load draws harmonic current and reactive power from the system leading the source current to be non-sinusoidal with THD of 7%. After integrating the SAPF into the system, the reactive power and harmonic current required for non-linear load can be compensated such that the source current becomes sinusoidal with reduced THD of 2.96%.